\documentclass[aip,reprint]{revtex4-1}
\usepackage{graphicx}% Include figure files
\draft % marks overfull lines with a black rule on the right
\usepackage{xcolor}
\usepackage{siunitx}
\usepackage{graphicx}

\begin{document}

\title{MBE-grown virtual substrates for quantum dots emitting in the telecom O- and C-bands}

%**********************************
\author{B. Scaparra}

\affiliation{Walter Schottky Institut, Technical University of Munich, Germany}
\affiliation{Technical University of Munich; TUM School of School of Computation, Information and Technology, Department of Electrical Engineering, Germany}
\affiliation{Munich Center for Quantum Science and Technology (MCQST), Germany}

%%%%%%%*****

\author{E. Sirotti}
\affiliation{Walter Schottky Institut, Technical University of Munich, Germany}
\affiliation{Technical University of Munich; TUM School of School of Natural Sciences, Department of Physics, Germany}

%%%**

\author{A. Ajay}

\affiliation{Walter Schottky Institut, Technical University of Munich, Germany}
\affiliation{Technical University of Munich; TUM School of School of Natural Sciences, Department of Physics, Germany}
\affiliation{Munich Center for Quantum Science and Technology (MCQST), Germany}

%%%%%%%%%****

\author{B. Jonas}

\affiliation{Walter Schottky Institut, Technical University of Munich, Germany}
\affiliation{Technical University of Munich; TUM School of School of Computation, Information and Technology, Department of Electrical Engineering, Germany}
\affiliation{Munich Center for Quantum Science and Technology (MCQST), Germany}

%%********

\author{B. Costa}

\affiliation{Walter Schottky Institut, Technical University of Munich, Germany}
\affiliation{Technical University of Munich; TUM School of School of Computation, Information and Technology, Department of Electrical Engineering, Germany}
\affiliation{Munich Center for Quantum Science and Technology (MCQST), Germany}

%%********

\author{H. Riedl}

\affiliation{Walter Schottky Institut, Technical University of Munich, Germany}
\affiliation{Technical University of Munich; TUM School of School of Natural Sciences, Department of Physics, Germany}
\affiliation{Munich Center for Quantum Science and Technology (MCQST), Germany}

%%*******
\author{P. Avdienko}

\affiliation{Walter Schottky Institut, Technical University of Munich, Germany}
\affiliation{Technical University of Munich; TUM School of School of Natural Sciences, Department of Physics, Germany}
\affiliation{Munich Center for Quantum Science and Technology (MCQST), Germany}

%****************

\author{I. D. Sharp}

\affiliation{Walter Schottky Institut, Technical University of Munich, Germany}
\affiliation{Technical University of Munich; TUM School of School of Natural Sciences, Department of Physics, Germany}

%%********
\author{G. Koblmüller}

\affiliation{Walter Schottky Institut, Technical University of Munich, Germany}
\affiliation{Technical University of Munich; TUM School of School of Natural Sciences, Department of Physics, Germany}
\affiliation{Munich Center for Quantum Science and Technology (MCQST), Germany}

%%%%%%*******
\author{E. Zallo}

\affiliation{Walter Schottky Institut, Technical University of Munich, Germany}
\affiliation{Technical University of Munich; TUM School of School of Natural Sciences, Department of Physics, Germany}
\affiliation{Munich Center for Quantum Science and Technology (MCQST), Germany}

%%********
\author{J. J. Finley}

\affiliation{Walter Schottky Institut, Technical University of Munich, Germany}
\affiliation{Technical University of Munich; TUM School of School of Natural Sciences, Department of Physics, Germany}
\affiliation{Munich Center for Quantum Science and Technology (MCQST), Germany}

%%********
\author{K. Müller}

\affiliation{Walter Schottky Institut, Technical University of Munich, Germany}
\affiliation{Technical University of Munich; TUM School of School of Computation, Information and Technology, Department of Electrical Engineering, Germany}
\affiliation{Munich Center for Quantum Science and Technology (MCQST), Germany}

\email[bianca.scaparra@wsi.tum.de]{}

\begin{abstract}

InAs semiconductor quantum dots (QDs) emitting in the near infrared are promising platforms for on-demand single-photon sources and spin-photon interfaces. However, the realization of quantum-photonic nanodevices emitting in the second and third telecom windows with similar performance remains an open challenge. Here, we report an optimized heterostructure design for QDs emitting in the O- and C-bands grown by means of molecular beam epitaxy. The InAs QDs are grown on compositionally graded InGaAs buffers, which act as virtual substrates, and are embedded in mostly relaxed active regions. Reciprocal space maps of the indium profiles and optical absorption spectra are used to optimize In$_{\text{0.22}}$Ga$_{\text{0.78}}$As and In$_{\text{0.30}}$Ga$_{\text{0.70}}$As active regions, accounting for the chosen indium grading profile. This approach results in a tunable QD photoluminescence (PL) emission from 1200 up to \SI{1600}{nm}. Power and polarization dependent micro-PL measurements performed at \SI{4}{K} reveal exciton-biexciton complexes from quantum dots emitting in the telecom O- and C-bands. The presented study establishes a flexible platform that can be an essential component for advanced photonic devices based on InAs/GaAs that serve as building blocks for future quantum networks.

\end{abstract}

\maketitle

\section{Introduction} 

\indent Photonic quantum technologies require sources that emit photons  at a fast rate and with a high degree of indistinguishability. Semiconductor quantum dots (QDs) emitting at \SI{950}{nm} or \SI{785}{nm} have been demonstrated to be very promising systems to meet these demands\cite{Senellart17, Hanschke18, Sbresny22}.
Indeed, the implementation of resonators, combined with the ability to tune the QDs emission wavelengths via the Stark effect and to electrically control the surrounding electronic environment, make InAs QDs excellent building blocks for quantum communication protocols\cite{Somaschi16, Kolatschek19, Gazzano22, Sapienza2015}. In this regard, the possibility to achieve equal performance at telecom wavelengths would be especially appealing due to the pre-existing fiber infrastructure and low propagation losses \cite{Gyger22}. Emission in the telecom C-band is obtained by means of InAs QDs based heterostructures grown on InP substrates \cite{Benyoucef13,Holewa24,Muller18}, whereas compositionally graded InGaAs layers are needed for the case of InAs QDs on GaAs substrates to reduce their lattice mismatch \cite{Paul17, Wroński21, Wyborski23}. While the InAs/InP system allows for direct implementation of cavities by using Bragg gratings or photonic crystals \cite{Nawrath23, Kaupp23, Phillips24, Jeong22}, an optimized nonlinear grading profile was suggested for the second heterostructure \cite{Sittig22}. However, in that work the relaxed portion of the graded layer was included in the fabricated resonator, thus limiting the further implementation of electrical contacts or more elaborate post-processing. GaAs substrates offer the possibility to grow lattice-matched high refractive index contrast distributed Bragg reflectors and are less brittle than InP substrates. Additionally, metamorphic laser heterostructures emitting in the telecom bands grown on GaAs substrates have proven to be compatible with the implementation in the active region of gate-tunable devices and sacrificial layers\cite{Kwoen21, Kwoen21_2, Liang24}. A similar approach has also been effective in realizing emission sources in the telecom C-band\cite{Semenova08}, however a tunable indium grading design resulting in high-quality QD emission from telecom O- to the C-bands is still coveted.\\
\indent In this paper, we develop a grading design in which the InAs QD layer is embedded in an InGaAs active region with fixed indium content that is carefully chosen depending on the maximum relaxation reached in the underlying graded layer. Using reciprocal space maps and optical absorption measurements, we determine the indium profiles that best match the used indium grading rate. In particular, we identify a favorable indium concentration step-back value between the matrix and graded layer as a function of the grading rate of the latter. The active region acts as an independent substrate with chosen lattice constant and the dislocations are mostly confined to the relaxed part of the graded layer. By embedding a single QD layer in the active region we demonstrate the tunability of the lattice constant of the final substrate resulting in a QD emission in both the telecom O- and C-bands. Power and polarization dependent photoluminescence measurements reveal bright and sharp lines with the typical exciton-biexciton behavior. This highlights the potential of the presented design for the realization of single-photon sources in the O- and C-bands.

\section{Methods}

\begin{figure}
 \includegraphics[scale=0.8,trim=0 0 0 0, clip]{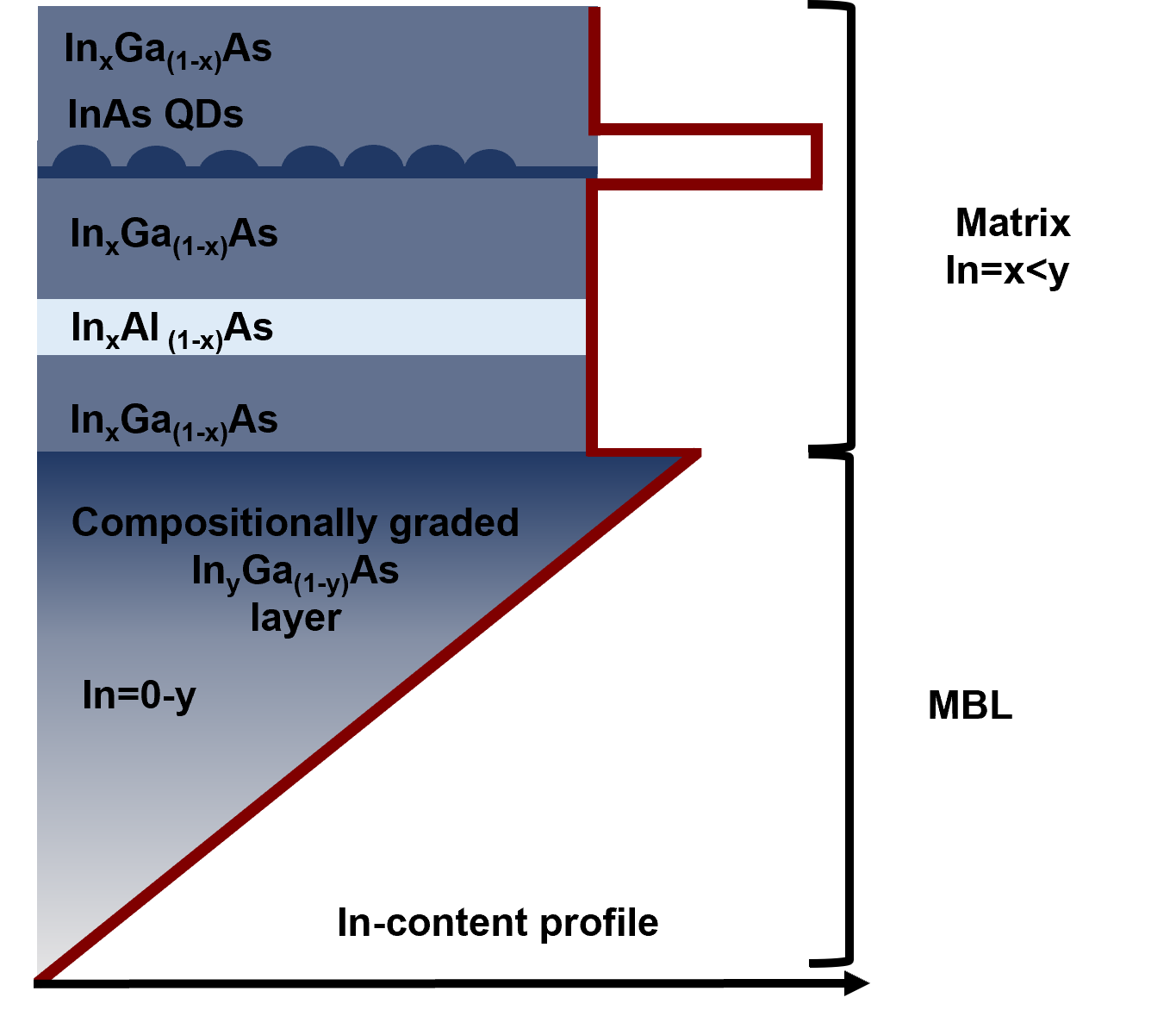}
\caption{\label{figure1} Schematic of the indium profile across the grown heterostructures. The InAs QDs and InAlAs layers are embedded in an InGaAs-matrix with fixed indium content grown on top of a compositionally graded InGaAs metamorphic buffer layer (MBL).}
\end{figure}

\indent The studied samples were grown with a solid source Veeco Gen II molecular beam epitaxy (MBE) system on undoped GaAs(001) substrates.  
The growth was monitored by reflection high energy electron diffraction and the native oxide desorption from the substrate was used to calibrate the surface temperature. The layer profile is presented in Fig.~\ref{figure1} as a function of the indium content. First, a \SI{200}{nm}-thick GaAs buffer layer was grown, followed by a \SI{100}{nm}-thick AlAs layer. Both layers were grown at \SI{610}{^{\circ}C}. Subsequently, \SI{30}{nm} of GaAs and a compositionally graded In$_{\text{x}}$Ga$_{\text{1-x}}$As metamorphic buffer layer (MBL) were grown at \SI{470}{^{\circ}C} with an arsenic beam equivalent pressure of 1.6×10$^{-5}$ mbar. The group-III elemental fluxes (gallium, indium, aluminium) were calibrated in equivalent growth rate units of {\text{\AA}/s}. During the growth of the graded In$_{\text{x}}$Ga$_{\text{1-x}}$As layer, the gallium growth rate was kept constant at \SI{1}{\text{\AA/}s}, while the indium cell temperature was increased with a nominal rate of \SI{0.02}{^{\circ}C/s}. %More details on the growth parameters can be found in Ref\cite{Scaparra_2023}. 
The indium growth rate was then increased from \SI{0.05}{\text{\AA/}s} to either \SI{0.58}{\text{\AA/}s} or \SI{0.75}{\text{\AA/}s} depending on the desired final lattice constant of the designed virtual substrate. The graded layer was then followed by an InGaAs matrix, or active region. The matrix presents a fixed indium content and was grown with a lower indium cell temperature, as depicted by the step-back in the indium profile shown in Fig.~\ref{figure1}. The indium contents of the InGaAs matrices were varied from 19$\%$ to 36$\%$ depending on the maximum indium content of the underlying MBL. During the growth of the InGaAs matrix, the substrate temperature was increased to \SI{500}{^{\circ}C} and the arsenic beam equivalent pressure was reduced to 1.1×10$^{-5}$ mbar. After a \SI{150}{nm}-thick InGaAs layer, a \SI{100}{nm}-thick InAlAs layer was grown to prevent further propagation of dislocations into the matrix. The InAs QD layer, consisting of 2.2 monolayers, was grown at \SI{470}{^{\circ}C} and was embedded in the middle of a \SI{300}{nm}-thick InGaAs layer. To achieve a QD density gradient, the substrate rotation was stopped halfway through the growth of the QD layer.\\
\indent Structural characterization was performed via high-resolution X-ray diffraction measurements acquired with a Rigaku SmartLab system equipped with a copper anode. The Cu$_{K\alpha1}$ emission line ($\lambda$= 1.54059 \AA) was filtered in the incident beam path with a Ge(220)x2 monochromator for high-resolution measurements. Reciprocal space maps (RSMs) around the asymmetric GaAs (422) reflection were used to analyze the crystalline properties of the grown heterostructures. The absorption coefficients of the InGaAs matrices grown on top of the MBLs were measured with a custom-made photothermal deflection spectroscopy (PDS) system. The sample was placed into a cuvette filled with perfluorohexane and illuminated at normal incidence with monochromatic light obtained with a monochromator placed after a Halogen lamp. The modulation frequency of the incident light was set at 9 Hz. The absorption was probed with a red laser diode directed parallel to the surface of the sample. A 2D detector connected to a lock-in amplifier was used to track the deflected probe laser beam. The PDS signal was then converted to an absorption coefficient with the layer thickness determined from scanning electron microscopy measurements. Temperature-dependent PL spectroscopy from ensembles of InAs QDs was performed under continuous wave non-resonant excitation at \SI{660}{nm} by using a helium flow cryostat operating in the 4-\SI{300}{K} temperature range. Micro-PL measurements were carried out at \SI{4}{K} using a confocal microscopy setup based on continuous wave excitation at \SI{780}{nm} and an apochromatic objective with a numerical aperture of 0.81. The detected signal was analyzed using a spectrometer with a focal distance of \SI{750}{mm} equipped with an InGaAs linear array detector, providing a resolution of 51 $\mu$eV at \SI{1550}{nm} and of 71 $\mu$eV at \SI{1310}{nm}. Polarization-resolved measurements were carried out by detecting the emitted signal as a function of the angle of a half-waveplate placed in front of a linear polarizer in the detection path.\\

\section{Results and discussion}

\begin{figure}
\includegraphics[scale=0.8, trim=5 0 0 0, clip]{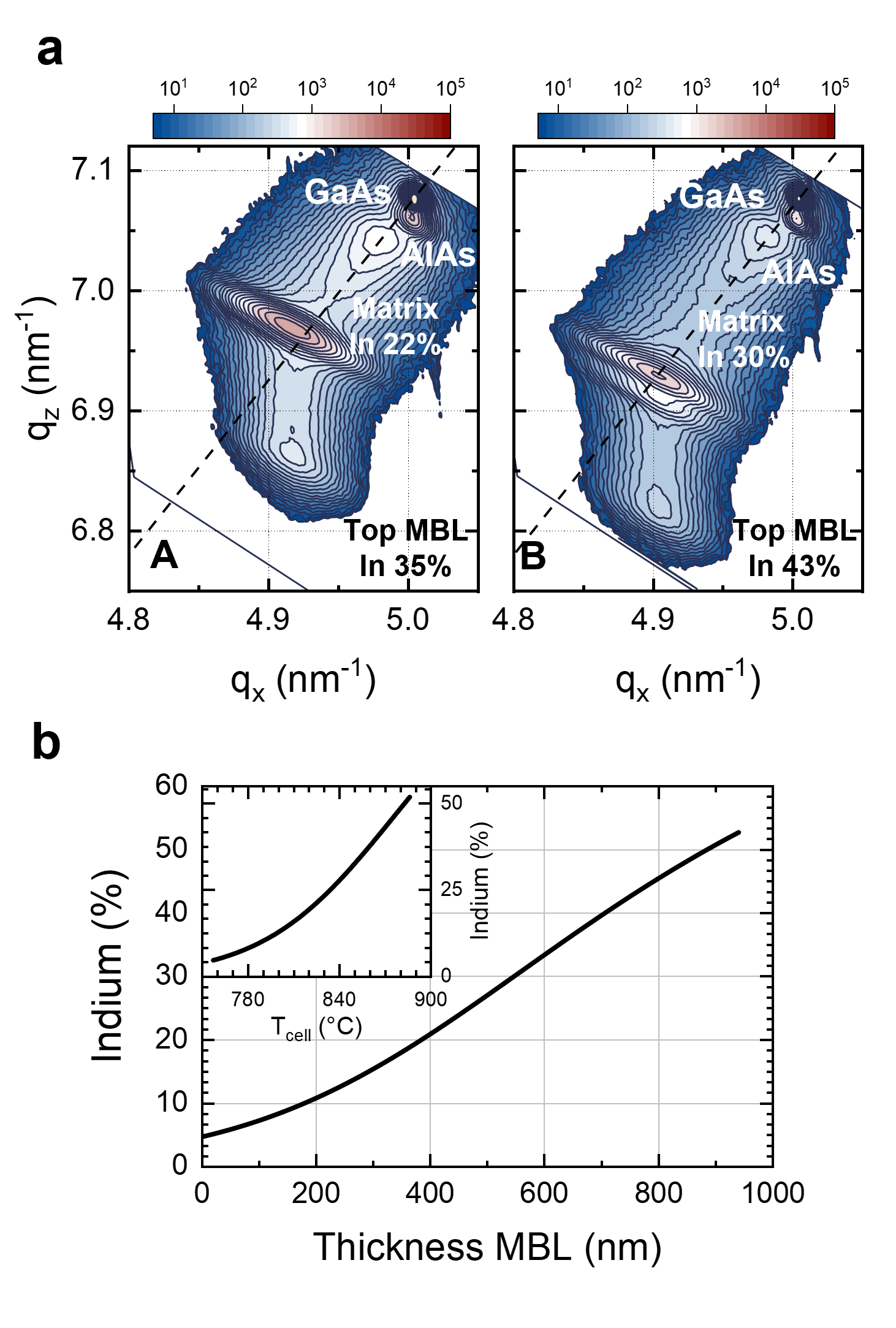}
 \caption{\label{figure2} \textbf{a} RSMs along the asymmetric (422) GaAs reflection from heterostructures with different maximum indium contents in the MBLs grown beneath InGaAs matrices with indium contents of 22$\%$ (sample \textbf{A}) and 30$\%$ (sample \textbf{B}). \textbf{b} Indium content as a function of the thickness of the InGaAs graded layer at a fixed gallium rate. The inset shows the dependence of the indium content as a function of the indium cell temperature T$_{cell}$ at a fixed gallium rate.}
 \end{figure}

\indent To realize the heterostructure presented in Fig.~\ref{figure1}, the compositionally graded In$_{\text{x}}$Ga$_{\text{1-x}}$As layer first had to be optimized in order to obtain the desired final lattice constant. Depending on its final value, the indium content of the matrix, or final substrate, was then selected. Fig.~\ref{figure2}\textbf{a} shows RSMs along the asymmetric (422) GaAs reflection of two different heterostructures, each consisting of a graded buffer layer and an InGaAs active region grown on top. The heterostructure \textbf{A} was grown with a maximum indium growth rate in the graded layer of \SI{0.58}{\text{\AA/}s}, while in sample \textbf{B} the maximum growth rate was \SI{0.75}{\text{\AA/}s}. The peak at larger coordinates arises from the GaAs substrate, while the diffraction signal spanning from the GaAs peak to the lowest q$_\text{x}$ and q$_\text{z}$ values originates from the graded MBL. Intensity maxima with coordinates close to the GaAs substrate correspond to areas of the MBL with lower indium content, whereas peaks located further away indicate regions with higher indium content. The dashed line illustrates the relaxation line, which corresponds to the direction in reciprocal space of a fully relaxed epitaxial layer. The signal arising from the layers with low indium content follows the relaxation line, revealing relaxed layers where the majority of the dislocations are confined \cite{Scaparra_2023}. Meanwhile, the part of the graded layers with higher indium content grows pseudomorphically to such relaxed layers, showing a constant q$_\text{x}$ value and indicating the indium content at which the MBL starts to show residual strain \cite{Tersoff93, Dunstan91}. The derived maximum indium contents in the graded layers are 35$\%$ (\textbf{A}) and 43$\%$ (\textbf{B}), determined following an approach similar to Ref. \cite{Scaparra_2023}. 
The signal spreading in diagonal directions stems from the mostly relaxed active regions with indium contents of 22$\%$ and 30$\%$, respectively. Both exhibit some mosaicity. As desired, the peaks arising from the layers in the active region have the same q$_\text{x}$ values as those of the relaxed part of the MBL at higher indium contents. Hence, the two layers show similar in-plane lattice constants and thus, the probability of dislocation propagation into the matrix is reduced \cite{Olsen96}. 
Due to the chosen grading rate and growth temperature of the InGaAs graded layers, we expect that near-equilibrium strain relaxation was reached \cite{Scaparra_2023}.\\
\indent It has been shown that the residual strain in the uppermost part of a linearly graded layer is independent of both its thickness and the maximum indium concentration \cite{Dunstan91,Tersoff93, Sacedon95, SOROKIN201683}. Thus, the indium step-back in the active region, needed to compensate for the residual strain of a graded layer, depends on the chosen grading rate\cite{Sacedon95}. Although the indium profile set by the temperature ramping rate of the indium cell is not strictly linear along the graded layer as shown in Fig.~\ref{figure2}\textbf{b}, we found that an indium step-back close to $\sim 13\%$ leads to a mostly relaxed active region for both sample \textbf{A} and \textbf{B}. This is consistent with previous literature, where for a linear grading rate of $\sim$30 In$\%$/$\mu$m, the indium step-back required to compensate for the residual strain was determined to be $\sim$8.3$\%$ \cite{Sacedon95, Dunstan91, Tersoff93}.
Consequently, the higher average grading rate used in this study ($\sim$49 In$\%$/$\mu$m) results in a larger degree of residual strain  \cite{Sacedon95}. Therefore, a larger indium step-back is required. The inset in Fig.~\ref{figure2}\textbf{b} shows the final indium profile with some non linearities across the graded layer. This is the result of the indium content dependence as a function of the cell temperature (see the inset at fixed gallium rate of \SI{1}{\text{\AA/}s}). However, similar values of residual strain are found for both samples (see Fig.~\ref{figure2}\textbf{a}) thus requiring the same indium step back value. Therefore, to minimize a further relaxation of the graded layer, the indium content selected for the matrix was closely matched to the value determined based on the chosen grading profile.\\ 
\begin{figure*}
\includegraphics[width=1\textwidth]{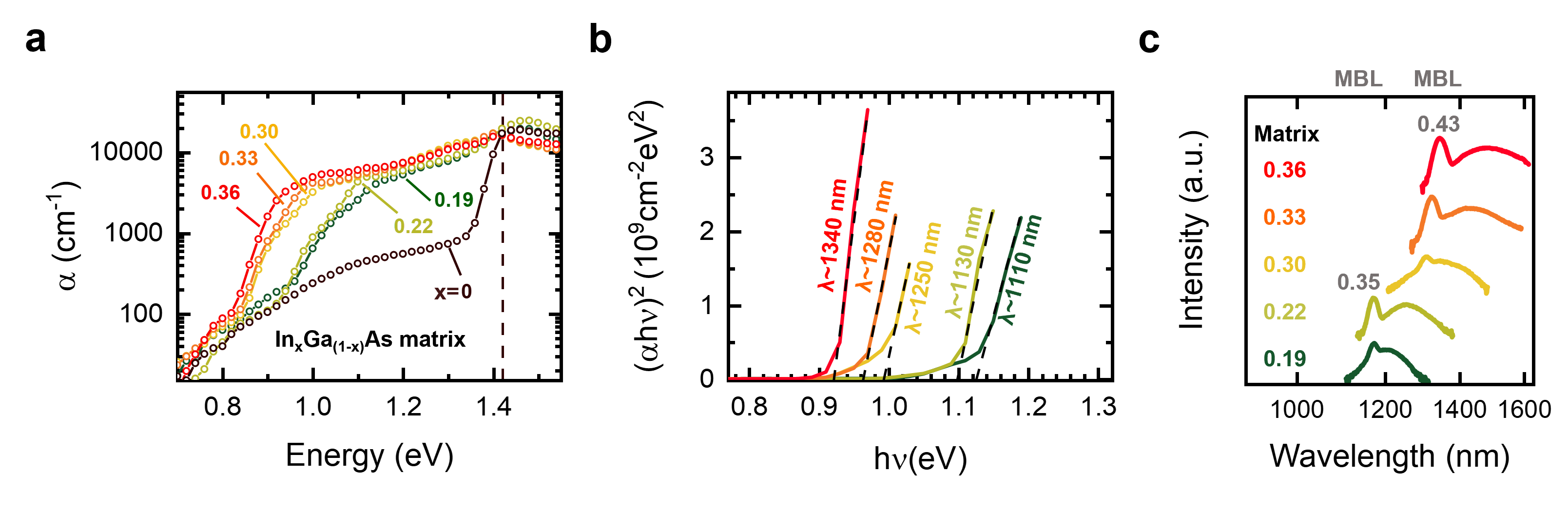}
 \caption{\label{figure3} \textbf{a} Absorption coefficients $\alpha$ recorded by PDS from samples with indium content in the matrices ranging between 19$\%$ and 36$\%$ (maximum indium content in the MBL of 35$\%$ and 43$\%$). \textbf {b} Tauc plot of ($\alpha$h$\nu$)$^{2}$ plotted versus the absorption energy h$\nu$. The intersections of the dashed lines with the energy axis indicate the direct absorption edges of the InGaAs matrices. The determined bandgap is labeled as $\lambda$. \textbf{c} Ensemble PL spectra recorded at \SI{10}{K} from QDs embedded in heterostructures with different indium content in the matrices grown on MBLs having 35$\%$ and 43$\%$ maximum indium content.}
 \end{figure*}
\indent Fig.~\ref{figure3}\textbf{a} shows optical absorption spectra measured via PDS \cite{Jackson81} from samples with active regions grown with different indium contents. Each curve is labeled with the indium content of the active region determined from asymmetric RSMs. All spectra exhibit similar absorption around \SI{1.42}{eV} (dashed line), which is attributed to the absorption at the bandgap of the GaAs substrate. In addition, absorption onsets around \SI{1.1}{eV} (\SI{0.9}{eV}) are observed for samples in which InGaAs matrices are grown atop MBLs with maximum indium contents of 35$\%$ (43$\%$). The linear regressions of the Tauc plots shown in Fig.~\ref{figure3}\textbf{b} allow to estimate the optical bandgaps of the InGaAs matrices. The color scale follows the indium contents labeled in Fig.~\ref{figure3}\textbf{a}. Prior to performing the linear regression, each spectrum was divided by the spectrum recorded from the bare GaAs substrate, as the optical absorption of the substrate is stronger than the one from the matrices. The intersections of the linear regions of the ($\alpha$h$\nu$)$^{2}$ vs.h$\nu$ plots with the energy axis indicate the optical bandgaps\cite{DOLGONOS201643}, as labeled in Fig.~\ref{figure3}\textbf{b} with $\lambda$. The optical bandgaps obtained by linear regression correspond to indium contents similar to the ones measured via RSMs.\\
\begin{figure*}
\includegraphics[width=0.7\textwidth]{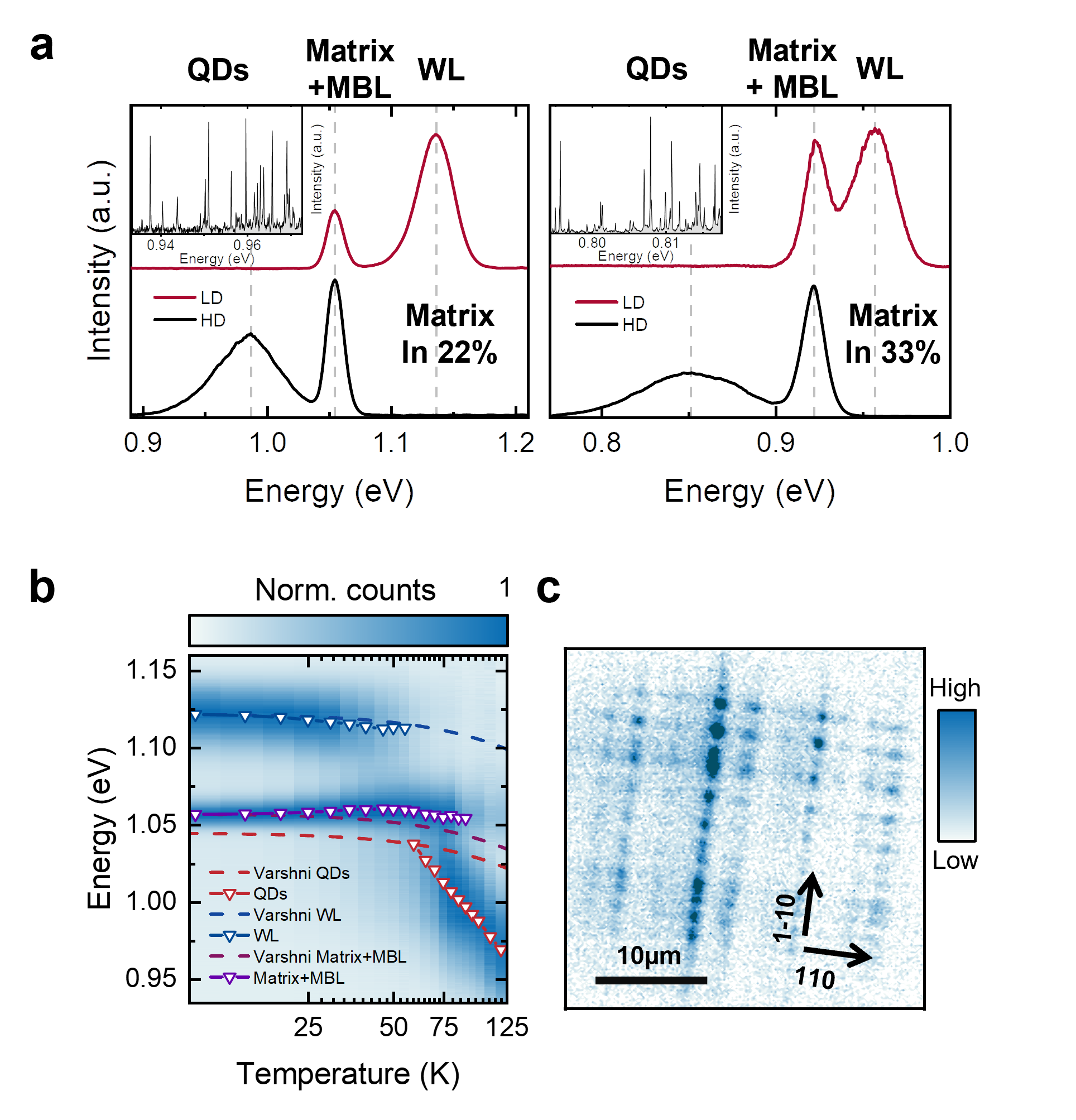}
 \caption{\label{figure4} \textbf{a} Representative PL spectra acquired in regions with low (LD, red) and high (HD, black) QD density. (Insets) Micro-photoluminescence spectra for each of the presented samples measured at \SI{4}{K}, showing single emission lines from a set of QDs. \textbf{b} Ensemble PL spectra as a function of the lattice temperature from heterostructure \textbf{A}. The peak positions of QD, matrix, and WL (triangles) are plotted together with the bandgap shrinkage of bulk InAs (dashed lines) shifted with repect to the QDs, matrix, and WL emissions at \SI{10}{K}. \textbf{c} Micro-PL map of sample \textbf{A} recorded at \SI{4}{K}, showing QDs emitting in the spectral range of 1250-\SI{1350}{nm}.}
\end{figure*}   
\indent To prove whether the presented sample designs lead to a QD emission in the telecom bands, ensemble PL measurements were taken at low temperature by exciting at \SI{780}{nm}. As shown in Fig.~\ref{figure3}\textbf{c}, the variation of the lattice mismatch due to different indium contents in the active region leads to a shift of the QD PL emission to the second and even further to the third telecom windows. Two peaks can be distinguished in each spectrum: the one at shorter wavelengths is attributed to charges recombining in the graded layer and the matrix, while the emission at longer wavelengths is attributed to the QDs. Samples with active regions with indium contents between 19$\%$ and 22$\%$ (maximum MBL indium content of 35$\%$) lead to a QD emission that can be tuned across the O-band. When the indium in the matrices ranges between 30 and 36$\%$ (maximum MBL indium content of 43$\%$) the QD emission shifts up to the telecom C-band. It is important to note that samples with indium step-back much lower than 13$\%$, e.g. the one with a indium content of 36$\%$, result in an active region that is compressively strained with respect to the matrix. Since this could lead to further relaxation in the graded layer\cite{Sacedon95}, such sample was not considered for further optical measurements.\\
\begin{figure*}
\includegraphics[width=1\textwidth]{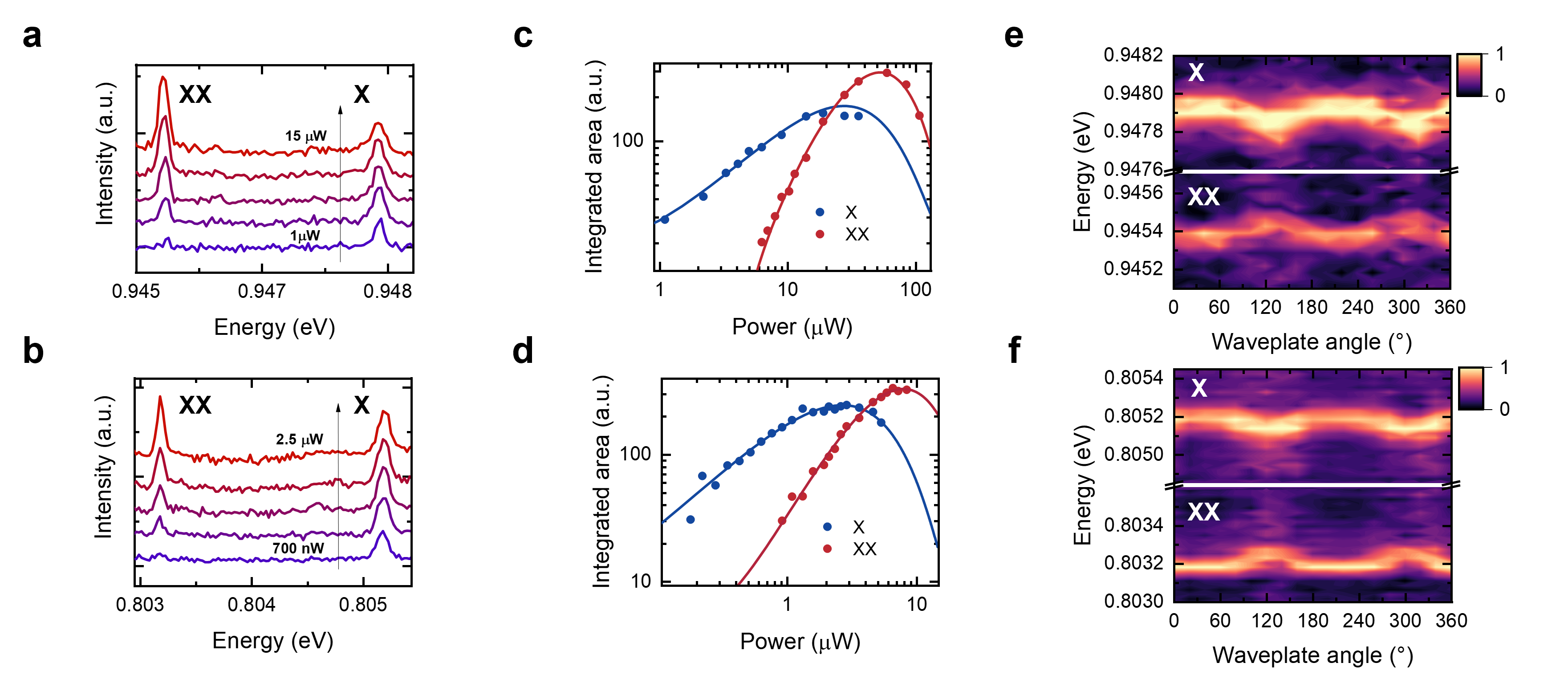}
\caption{\label{figure5} \textbf{a}-\textbf{b} Micro-PL spectra of QDs emitting in the O- and C-band measured from samples \textbf{A} and \textbf{B}. The spectra were recorded at \SI{4}{K} under non-resonant excitation at different excitation powers. \textbf{c}-\textbf{d} Integrated intensity of the emission lines shown in \textbf{a}-\textbf{b} respectively. The data were fitted with Poissonian distributions I(P)= a$\cdot$(P$^n$/n!)$\cdot$e$^{b\cdot P}$+I$_0$ as a function of the power, yielding the probability that the QD is occupied with n=1 (blue) or n=2 (red) excitons complexes. \textbf{e}-\textbf{f} Polarization dependence of the emission energies of the X and XX emission lines from QDs emitting in the telecom O- and C-bands respectively.}
\end{figure*}
\indent Fig.~\ref{figure4}\textbf{a} shows ensemble spectra obtained under above-bandgap excitation at a power of 600 $\mu$W from samples with matrix indium contents of 22 and 33$\%$, recorded in regions of the wafer with different QD densities. If the QD density is low enough (LD, red), the injected charges mostly recombine in the wetting layer (WL), leading to the appearance of the peaks  at higher energies. Meanwhile, in regions where the QD density increases (HD, black), the signal from charges recombining in the WL is quenched and the QD peak appears at lower energies in both graphs. The peaks centered around \SI{1.05}{eV} and \SI{0.92}{eV} are due to charges recombining in the matrix and in the uppermost regions of the MBL (see Fig.~\ref{figure1}). The insets show bright and sharp lines in the telecom O and C-bands attributed to PL emission of a few QDs excited at a wavelength of \SI{780}{nm}. The respective full width at half maximum (FWHM) values of the emissions is 89 $\pm$15 $\mu$eV (matrix 22$\%$) and 90 $\pm$22 $\mu$eV (matrix 33$\%$), which are limited by the spectrometer resolution. \\
\indent The assignment of the peaks presented in Fig.~\ref{figure4}\textbf{a} is further confirmed by the temperature dependent PL recorded from sample \textbf{A}, as shown in Fig.~\ref{figure4}\textbf{b}. The spectra were recorded in a similar region where the spectrum LD was measured and were normalized with respect to the peak maxima. The measurements were performed under above-bandgap excitation with a power of 300 $\mu$W. Each spectrum was fitted with Gaussian profiles and the corresponding center peak positions are shown for each temperature. The fitted spectral positions are overlaid with the InAs Varshni relation (dashed lines), which indicates the dependence of the bandgap of bulk InAs on the lattice temperature \cite{VARSHNI67, Vurgaftman01}. At \SI{10}{K} the emissions attributed to the WL (blue triangles) and the matrix (purple triangles) are visible. When the temperature reaches \SI{60}{K}, another peak at lower energies appears, which is attributed to the QD emission (red triangles), while the one that stems from the WL quenches. This marks the temperature at which the charge transfer mechanism from the WL states to the QD ones becomes favorable \cite{Syperek13, Scaparra_2023}. The QD peaks show the typical steeper spectral redshift when compared to the Varshni relation (dashed lines). This can be attributed to charge redistribution into QDs with lower energy ground states \cite{Wyborski22, Scaparra_2023}.
In contrast, the WL peak follows the behavior given by the Varshni relation until it quenches at \SI{60}{K} in favor of the QD emission. Meanwhile, the signal arising from the matrix and graded layer persists up to \SI{110}{K} before vanishing. Fig.~\ref{figure4}\textbf{c} shows a 25x25 $\mu$m$^{2}$ micro-PL spatial map from sample \textbf{A} recorded at \SI{4}{K}. The existence of periodic modulations on the surface, and thus in the QD formation, is a hallmark of the underlying dislocation network due to the plastic relaxation in the MBL\cite{Semenova08, Paul17, Scaparra_2023}. The image shows an area of the sample in which the QD density is on the order of $\sim$ 10$^{9}$ cm$^{-2}$. The pattern resulting from optically active QDs resembles the one from the cross-hatched surfaces along the [110]-[1$\bar{1}$0] directions typical for metamorphic surfaces \cite{Cordier00, Cordier03, Romanato99, jung17}. As implied by the figure, the QD nucleation tends to follow the surface undulations along the [1-10] direction\cite{Scaparra_2023, Hausler96, Paul17} and it is affected by the indium composition modulation and thus, strain fluctuation in the active region\cite{Semenova08}.\\
\indent Micro-PL experiments on samples \textbf{A} and \textbf{B} were carried out at \SI{4}{K} in order to identify excitonic complexes in the second and third telecom windows. Figs.~\ref{figure5}\textbf{a-b} show the power dependent PL of exciton-biexciton complexes. The marked emission lines can be identified as excitons (X) and biexcitons (XX) separated by binding energies of 2.5 and \SI{2}{meV} in the O and C-telecom bands, respectively, similar to the values reported in literature \cite{Wyborski23, Wyborski22, Scaparra_2023}. This assignment is further corroborated via power dependent and polarization resolved measurements. In Figs.~\ref{figure5}\textbf{c-d} the integrated areas of the transition are presented as a function of the excitation power. The data follow Poissonian distributions, where an average number of generated excitons equal to one (blue dots) correspond to X transitions. Meanwhile, the data corresponding to an average number of two generated excitons (red dots) correspond to XX transitions\cite{Abbarchi09}. At low powers, the X transitions show a linear dependence on the excitation power, while the XX transitions only appear at higher powers and exhibit a quadratic behavior. For powers greater than the ones at which the X emission saturates, the XX emission prevails. The FWHMs from X and XX of Fig.~\ref{figure5}\textbf{a}, measured in the X saturation regime, are 123 and 90 $\mu$eV, respectively, while values of 107 and 54 $\mu$eV are measured from the respective lines in Figs.~\ref{figure5}\textbf{b}. These values are lower than the average ones reported for QDs grown directly on MBE-grown MBLs (around 200-300 $\mu$eV for emissions in the telecom O- and C-bands)\cite{Wyborski23}. Polarization dependent PL measurements shown in Fig.~\ref{figure5}\textbf{e-f} support the assigned excitonic behaviors. The measurements were carried out in the saturation regime of the X transitions. The counter-oscillating behavior of the spectral oscillations that present the same amplitude is typical of a XX-X cascade \cite{Young05}. From fits to these data, we extract fine structure splittings of 60$\pm$6 $\mu$eV and 55$\pm$6 $\mu$eV for the QDs shown in Figs.~\ref{figure5}\textbf{e-f}, respectively. These results confirm the effectiveness of the studied indium profiles along the heterostructure in tuning the QD emission over a large spectral region. They also demonstrate the successful design of the virtual substrates, which can be overgrown with substrates with tailored lattice constants.  

\section{Conclusions}

\indent In this paper, we investigated indium concentration profiles within metamorphic substrates tailored to achieve a QD emission in the telecom O- and C-bands. The graded layers were overgrown with InGaAs matrices and act as virtual substrates. By means of reciprocal space maps and optical absorption measurements, the indium step-back value between MBL and active region was optimized to guarantee a good relaxation degree of the latter for the adopted indium grading profile. Low temperature PL spectra showed the tunability of the QD emission as a function of the indium content in the matrix. The reported excitonic complexes in the telecom O- and C-bands proved the effectiveness of the heterostructure design also showing narrow linewidths of the excitonic transitions. One major advantage of the presented heterostructures is the possibility to grow InAs QDs on optimized substrates with tailored lattice constants. The control of the mismatch between the QDs and the final substrate enables the tuning of their emission from 1200 up to \SI{1600}{nm}. Another advantage lies in the confinement of dislocations  caused by the plastic relaxation of the lattice constant, primarily within the graded layer grown beneath the final substrate. The possibility of growing barriers or superlattices in the active region beneath the QDs helps prevent further propagation of the dislocations across the matrix. We anticipate this study to be the first step towards developing quantum light emitting diodes at \SI{1550}{nm} and defect-free substrates for integrated photonics on silicon based on GaAs/InAs material systems.

\begin{acknowledgments}

We gratefully acknowledge  financial support from the German Federal Ministry of Education and Research via the funding program Photonics Research Germany (Contract No. 13N14846), the European Union's Horizon 2020 research and innovation program under Grants Agreement No. 862035 (QLUSTER), the Deutsche Forschungsgemeinschaft (DFG, German Research Foundation) via the projects MQCL (INST 95/1220-1), CNLG (MU 4215$/$4-1) and Germany's Excellence Strategy (MCQST, EXC-2111, 390814868), the Bavarian State Ministry of Science and Arts (StMWK) via the project EQAP and the Bavarian Ministry of Economic Affairs (StMWi) via project 6GQT.\\
G.K. acknowledges financial support by the ERC project QUANtIC (ID: 771747) funded by the European Research Council.\\

\end{acknowledgments}

\section*{references}
\bibliography{aipsamp}

\end{document}